\documentclass[12pt]{article}
\usepackage{amsmath,amssymb,amsthm,amsxtra,overpic,bbm,bm,epsfig}
\renewcommand{\baselinestretch}{1.2}
\def\thefootnote{\fnsymbol{footnote}}

\begin{document}

\vspace{0.2cm}

\begin{center}
{\Large\bf Interference effects in reactor antineutrino oscillations}
\end{center}

\vspace{0.1cm}

\begin{center}
{\bf Zhi-zhong Xing}$^{1),~2)}$
\footnote{E-mail: xingzz@ihep.ac.cn} \\
$1)$ {Institute of High Energy Physics, and School of Physical Sciences, \\
\hspace{0.3cm} University of Chinese Academy of
Sciences, Beijing 100049, China \\
$2)$ Center for High Energy Physics, Peking University, Beijing 100080, China}
\end{center}

\vspace{1.5cm}

\begin{abstract}
We reformulate the probabilities of disappearance neutrino or antineutrino
oscillations so as to single out the
interference term proportional to the product of
$\sin[\Delta m^2_{21} L/(4 E)]$ and
$\sin[(\Delta m^2_{31} + \Delta m^2_{32}) L/(4 E)]$,
which is transparently sensitive to the neutrino mass ordering.
We elaborate this issue for a reactor-based antineutrino
oscillation experiment like JUNO, and take account of terrestrial matter
effects. If a light sterile neutrino species contributes to
${\cal P}(\overline{\nu}^{}_e \to \overline{\nu}^{}_e)$,
we find that there will be a new interference
term proportional to the product of $\sin^2 2\theta^{}_{14}$,
$\sin[\Delta m^2_{21} L/(4 E)]$ and
$\sin[(\Delta m^2_{41} + \Delta m^2_{42}) L/
(4 E)]$ in the standard parametrization of the $(3+1)\times (3+1)$
active-sterile neutrino mixing matrix.
\end{abstract}

\begin{flushleft}
\hspace{0.8cm} PACS number(s): 14.60.Pq, 13.10.+q, 25.30.Pt \\
\hspace{0.8cm} Keywords: neutrino mass ordering,
reactor antineutrino oscillation, sterile neutrino
\end{flushleft}

\def\thefootnote{\arabic{footnote}}
\setcounter{footnote}{0}

\newpage

\section{Motivation}

Reactors have been playing an important role in the development of experimental neutrino physics: from the discovery of the electron antineutrino --- Wolfgang Pauli's
hypothetical particle carrying away a part of the energy emitted from the
beta decay \cite{Reines},
to the discoveries of $\overline{\nu}^{}_e \to \overline{\nu}^{}_e$ oscillations
at two different baseline scales \cite{KamLAND,DYB}. Today the reactor antineutrino
oscillations remain a powerful tool to probe the neutrino mass ordering
and search for possible new physics beyond the standard three-flavor scheme,
such as the existence of light or heavy sterile neutrinos \cite{JUNO}.

It is actually the interference effect between the oscillation terms driven by the
neutrino mass-squared differences $\Delta m^2_{21} \equiv m^2_2 - m^2_1$
and $\Delta m^2_{3i} \equiv m^2_3 - m^2_i$
(for $i = 1, 2$) that allows us to detect the unknown sign of
$\Delta m^2_{3i}$ in a reactor-based medium-baseline antineutrino oscillation
experiment like JUNO \cite{Wang}. Given the very fact of $|\Delta m^2_{31}|
\simeq |\Delta m^2_{32}| \sim 30 \Delta m^2_{21}$ with
$\Delta m^2_{21} >0$ \cite{Lisi}, the normal mass ordering $m^{}_1 < m^{}_2 < m^{}_3$
and the inverted one $m^{}_3 < m^{}_1 < m^{}_2$ correspond to
$\Delta m^2_{3i} >0$ and $\Delta m^2_{3i} <0$ (for $i = 1, 2$), respectively. Namely,
$\Delta m^2_{31}$ and $\Delta m^2_{32}$ must be of the same sign.

Since the JUNO experiment has optimized its baseline to be
about $55 ~{\rm km}$, it is aimed to measure the $\Delta m^2_{3i}$-caused
fine structure in the energy spectrum of $\overline{\nu}^{}_e \to \overline{\nu}^{}_e$
oscillations driven by $\Delta m^2_{21}$, which can be used to
discriminate between the normal and inverted neutrino mass orderings.
In this case the energy resolution of the JUNO detector is required
to be extraordinarily good \cite{Li2013}.

Note that in such a precision measurement the probability of
$\overline{\nu}^{}_e \to \overline{\nu}^{}_e$ oscillations depends on all the
three neutrino mass-squared differences, but only two of them are independent
because of the linear correlation $\Delta m^2_{21} = \Delta m^2_{31} - \Delta m^2_{32}$.
Hence the interference term in the oscillation probability
${\cal P}(\overline{\nu}^{}_e \to \overline{\nu}^{}_e)$
is {\it analytically} not unique \cite{Petcov}, depending on which of the
following combinations you
like to choose: $\Delta m^2_{21}$ and $\Delta m^2_{31}$ \cite{Ge}, or
$\Delta m^2_{21}$ and $\Delta m^2_{32}$ \cite{Ge2}, or
$\Delta m^2_{31}$ and $\Delta m^2_{32}$, or
$\Delta m^2_{21}$ and $\Delta m^2_{31} + \Delta m^2_{32}$
\cite{Cahn2013,Lisi2014,Wang2016,Li2016,Zhu2018}, etc.

In this short note we argue that the favorite {\it analytical} expression
of ${\cal P}(\overline{\nu}^{}_e \to \overline{\nu}^{}_e)$ should make the
interference effect most transparent, especially its clear reflection of the
neutrino mass ordering. In this sense the combination of
$\Delta m^2_{21}$ and $\Delta m^2_{31} + \Delta m^2_{32}$ turns out
to be more favored than the others
\footnote{At this point it is worth pointing out that
the definitions $\delta m^2 \equiv m^2_2 - m^2_1$ and
$\Delta m^2 \equiv m^2_3 - \left(m^2_1 + m^2_2\right)/2$
by Fogli and Lisi in their global analysis of neutrino oscillation
data \cite{Fogli} are equivalent to what we are recommending here,
simply because $\delta m^2 = \Delta m^2_{21}$ and
$\Delta m^2 = \left(\Delta m^2_{31} + \Delta m^2_{32}\right)/2$ hold.}.
The same point of view is also applicable to the case in
which one or more light sterile neutrinos participate in the reactor
antineutrino oscillations.

To be more general, we shall first reformulate the probabilities of
disappearance-type neutrino oscillations to single out the
interference term proportional to the product of
$\sin[\Delta m^2_{21} L/(4 E)]$ and
$\sin[(\Delta m^2_{31} + \Delta m^2_{32}) L/(4 E)]$,
which is sensitive to the neutrino mass ordering in a transparent way.
Then we are going to focus on the $\overline{\nu}^{}_e \to \overline{\nu}^{}_e$
oscillations in a reactor-based medium-baseline experiment.
If a light sterile neutrino species is assumed to take part in the
reactor antineutrino oscillations, we find that there will be an extra
interference term proportional to the product of $\sin^2 2\theta^{}_{14}$,
$\sin[\Delta m^2_{21} L/(4 E)]$ and
$\sin[(\Delta m^2_{41} + \Delta m^2_{42}) L/
(4 E)]$, where $\theta^{}_{i4}$ and $\Delta m^2_{4i}
\equiv m^2_4 - m^2_i$ (for $i=1,2,3$) stand respectively for the
active-sterile neutrino mixing angles and mass-squared differences.
Whether such a new interference term could contaminate the standard
one or not depends on the possibly allowed ranges of those new parameters.

\section{In the standard three-flavor case}

In the standard three-flavor scheme the formula for the probabilities
of disappearance-type $\nu^{}_\alpha \to \nu^{}_\alpha$ and
$\overline{\nu}^{}_\alpha \to \overline{\nu}^{}_\alpha$ oscillations
(for $\alpha = e, \mu, \tau$) is well known. But here let us reformulate it
by taking account of the observation that the difference between
$\Delta m^2_{31}$ and $\Delta m^2_{32}$ has been fixed and their sum
is a signature of the neutrino mass ordering. The result is
\begin{eqnarray}
{\cal P}(\overline{\nu}^{}_\alpha \to \overline{\nu}^{}_\alpha)
& \hspace{-0.2cm} =  \hspace{-0.2cm} &
{\cal P}(\nu^{}_\alpha \to \nu^{}_\alpha) =
1 - 4\sum^3_{i<j} \left[|U^{}_{\alpha i}|^2 |U^{}_{\alpha j}|^2
\sin^2\frac{\Delta m^2_{ji} L}{4 E} \right]
\nonumber \\
& \hspace{-0.2cm} =  \hspace{-0.2cm} &
1 - 4 |U^{}_{\alpha 1}|^2 |U^{}_{\alpha 2}|^2
\sin^2 \frac{\Delta m^{2}_{21} L}{4 E}
- \hspace{0.1cm} 2 |U^{}_{\alpha 3}|^2 \left(1 - |U^{}_{\alpha 3}|^2\right)
\left(\sin^2 \frac{\Delta m^{2}_{31} L}{4 E}
+ \sin^2 \frac{\Delta m^{2}_{32} L}{4 E}\right)
\nonumber \\
& & \hspace{0.3cm} - \hspace{0.1cm} 2
|U^{}_{\alpha 3}|^2 \left(|U^{}_{\alpha 1}|^2 - |U^{}_{\alpha 2}|^2
\right) \sin \frac{\Delta m^{2}_{21} L}{4 E}
\sin \frac{\left(\Delta m^{2}_{31} + \Delta m^2_{32}\right) L}{4 E} \; ,
\end{eqnarray}
where $U^{}_{\alpha i}$ (for $i=1,2,3$) denote the elements of the
$3\times 3$ Pontecorvo-Maki-Nakagawa-Sakata (PMNS) \cite{PMNS}
flavor mixing matrix $U$. This new formula makes it very transparent that
the last term describes the interference effect between the
oscillations driven by $\Delta m^2_{21} = \Delta m^2_{31} -
\Delta m^2_{32}$ and $\Delta m^2_{31} + \Delta m^2_{32}$, which is
therefore sensitive to the neutrino mass ordering.
In the standard parametrization of $U$ we have $|U^{}_{e1}| =
\cos\theta^{}_{12}\cos\theta^{}_{13}$,
$|U^{}_{e2}| = \sin\theta^{}_{12}\cos\theta^{}_{13}$ and
$|U^{}_{e3}| = \sin\theta^{}_{13}$. The explicit formula for
a reactor-based antineutrino oscillation experiment turns out to be
\begin{eqnarray}
{\cal P}(\overline{\nu}^{}_e \to \overline{\nu}^{}_e)
& \hspace{-0.2cm} =  \hspace{-0.2cm} &
1 - \hspace{0.05cm} \sin^2 2\theta^{}_{12} \cos^4\theta^{}_{13}
\sin^2 \frac{\Delta m^{2}_{21} L}{4 E} - \frac{1}{2}
\sin^2 2\theta^{}_{13} \left(\sin^2 \frac{\Delta m^{2}_{31} L}{4 E}
+ \sin^2 \frac{\Delta m^{2}_{32} L}{4 E}\right)   \hspace{0.5cm}
\nonumber \\
& & \hspace{0.3cm} - \hspace{0.1cm}
\frac{1}{2} \cos 2\theta^{}_{12} \sin^2 2\theta^{}_{13}
\sin \frac{\Delta m^{2}_{21} L}{4 E}
\sin \frac{\left(\Delta m^{2}_{31} + \Delta m^2_{32}\right) L}{4 E} \; .
\end{eqnarray}
Some straightforward remarks are in order.
\begin{itemize}
\item     The first oscillatory term on the right-hand side of Eq. (2) may
correspond to the KamLAND reactor antineutrino oscillation
experiment with an average baseline $L \sim 180$ km \cite{KamLAND}.

\item    The second oscillatory term in Eq. (2) is apparently responsible for
the short-baseline ($L \lesssim 2$ km) Daya Bay \cite{DYB}, RENO \cite{RENO}
and Double Chooz \cite{DC} experiments which are uniquely sensitive to the smallest
neutrino mixing angle $\theta^{}_{13}$.

\item    The last oscillatory term describes the fine interference effect
in ${\cal P}(\overline{\nu}^{}_e \to \overline{\nu}^{}_e)$ which will
be probed in the medium-baseline JUNO experiment with $L \simeq 55$ km
\cite{JUNO}. It is proportional to
$\sin[(\Delta m^2_{31} + \Delta m^2_{32}) L/(4 E)]$, and that is
why it is sensitive to the common unknown sign of $\Delta m^2_{31}$
and $\Delta m^2_{32}$. This term will cause a fine structure in the
energy spectrum of $\overline{\nu}^{}_e \to \overline{\nu}^{}_e$
oscillations driven by $\Delta m^2_{21}$, and hence the energy resolution
of the JUNO detector must be good enough to measure it.
\end{itemize}
It is also worth pointing out that the interference term in Eq. (2)
would disappear if $\theta^{}_{12} = 45^\circ$ or $\theta^{}_{13} =0^\circ$
held. Fortunately, neither of them is true because $\theta^{}_{12}
\simeq 34^\circ$ and $\theta^{}_{13} \simeq 8.5^\circ$ have been
well determined from current neutrino oscillation data \cite{Lisi}. Note that
$\theta^{}_{12} = 45^\circ$ would imply equal contributions of the
mass eigenstates $|\nu^{}_1\rangle$ and $|\nu^{}_2\rangle$ to the
flavor eigenstate $|\nu^{}_e\rangle$
\footnote{In the two-flavor neutrino oscillation scheme the mysterious
possibility $\theta^{}_{12} = 45^\circ$ would
make the matter-induced Mikheyev-Smirnov-Wolfenstein (MSW) resonance
effect \cite{MSW} impossible to show up.},
while $\theta^{}_{13} = 0^\circ$
would make the mass eigenstate $|\nu^{}_3\rangle$ absent or decoupled from
$|\nu^{}_e\rangle$.

Note that it is sometimes necessary to take into account terrestrial matter
effects in such a precision measurement of the reactor-based
$\overline{\nu}^{}_e \to \overline{\nu}^{}_e$
oscillations \cite{Lisi2014,Wang2016,Li2016,Zhu2018}. One may express
the matter-corrected oscillation probability
$\widetilde{\cal P}(\overline{\nu}^{}_e \to \overline{\nu}^{}_e)$ in
the same form as that of
${\cal P}(\overline{\nu}^{}_e \to \overline{\nu}^{}_e)$ in Eq. (2),
in terms of the corresponding {\it effective} neutrino mixing angles
$\widetilde{\theta}^{}_{ij}$ and mass-squared differences
$\Delta \widetilde{m}^2_{ij}$ in matter. In this case
the analytical approximations made in Ref. \cite{Zhu2016} lead us
to the following simple but instructive relations:
\begin{eqnarray}
\Delta \widetilde{m}^2_{21} & \hspace{-0.2cm} \simeq  \hspace{-0.2cm} &
\Delta m^2_{21} + A \cos 2\theta^{}_{12} \; ,
\nonumber \\
\Delta \widetilde{m}^2_{31} & \hspace{-0.2cm} \simeq  \hspace{-0.2cm} &
\Delta m^2_{31} + \frac{1}{2} A
\left(1 + \cos 2\theta^{}_{12}\right) \; ,
\nonumber \\
\Delta \widetilde{m}^2_{32} & \hspace{-0.2cm} \simeq  \hspace{-0.2cm} &
\Delta m^2_{32} + \frac{1}{2} A
\left(1 - \cos 2\theta^{}_{12}\right) \; ,
\end{eqnarray}
and $\Delta \widetilde{m}^2_{31} + \Delta \widetilde{m}^2_{32}
\simeq \Delta {m}^2_{31} + \Delta {m}^2_{32} + A$; together with
$\widetilde{\theta}^{}_{13} \simeq \theta^{}_{13}$ and
\begin{eqnarray}
\sin^2 2\widetilde{\theta}^{}_{12} & \hspace{-0.2cm} \simeq  \hspace{-0.2cm} &
\sin^2 2\theta^{}_{12} \left(1 - \frac{2 A}{\Delta m^2_{21}}
\cos 2\theta^{}_{12} \right) \; ,
\nonumber \\
\cos 2\widetilde{\theta}^{}_{12} & \hspace{-0.2cm} \simeq  \hspace{-0.2cm} &
\cos 2\theta^{}_{12} + \frac{A}{\Delta m^2_{21}} \sin^2 2\theta^{}_{12} \; ,
\end{eqnarray}
where $A = 2\sqrt{2} ~ G^{}_{\rm F} N^{}_e E$ is the matter
parameter and $A/\Delta m^2_{21} \simeq 1.05 \times 10^{-2} \times
E/\left(4 ~{\rm MeV}\right) \times 7.5 \times 10^{-5} ~ {\rm eV}^2
/\Delta m^2_{21}$ by taking $\rho \simeq 2.6 ~{\rm g}/{\rm cm}^3$
as a typical matter density of the Earth's crust \cite{Li2013}
\footnote{Here we would like to thank Jing-yu Zhu for kindly pointing out
two minor typing errors in Eqs. (10) and (18) of Ref. \cite{Li2013},
where $\sin^2 \theta^{}_{12}$ should be $\sin^2 2\theta^{}_{12}$.
The numerical results over there are not affected by these two errors.}.
Hence the terrestrial matter effects can reach the $1\%$ level.

\section{The (3+1) active-sterile neutrino mixing}

If one or more sterile neutrino species are assumed to exist and
slightly mix with the three active neutrinos, the
conventional $3\times 3$ PMNS matrix will just be
the submatrix of a $(3+n) \times (3+n)$ unitary matrix which describes the
whole flavor mixing effects among 3 active and $n$ sterile neutrinos:
\begin{eqnarray}
\left(\begin{matrix} \nu^{}_e \cr \nu^{}_\mu \cr \nu^{}_\tau \cr \vdots \cr
\end{matrix}
\right) = \left(\begin{matrix} U^{}_{e1} & U^{}_{e2} & U^{}_{e3} & \cdots \cr
U^{}_{\mu 1} & U^{}_{\mu 2} & U^{}_{\mu 3} & \cdots \cr
U^{}_{\tau 1} & U^{}_{\tau 2} & U^{}_{\tau 3} & \cdots \cr
\vdots & \vdots & \vdots & \ddots \cr \end{matrix} \right)
\left(\begin{matrix} \nu^{}_1 \cr \nu^{}_2 \cr \nu^{}_3 \cr \vdots \cr
\end{matrix}
\right) \; .
\end{eqnarray}
Depending on whether the sterile neutrinos participate in flavor oscillations
of three active neutrinos or not, they may give rise to direct or indirect
non-unitary effects in a realistic reactor antineutrino oscillation
experiment \cite{Xing2013}. For simplicity, here we only consider $n = 1$ for
a light sterile neutrino species which takes part in $\overline{\nu}^{}_e
\to \overline{\nu}^{}_e$ oscillations. In this case the disappearance
oscillation probability can be expressed as
\begin{eqnarray}
{\cal P}(\overline{\nu}^{}_e \to \overline{\nu}^{}_e)
& \hspace{-0.2cm} =  \hspace{-0.2cm} &
1 - 4\sum^3_{i<j} \left[|U^{}_{e i}|^2 |U^{}_{e j}|^2
\sin^2\frac{\Delta m^2_{ji} L}{4 E} \right]
- 4\sum^3_{i=1} \left[|U^{}_{e i}|^2 |U^{}_{e 4}|^2
\sin^2\frac{\Delta m^2_{4i} L}{4 E} \right]
\nonumber \\
& \hspace{-0.2cm} =  \hspace{-0.2cm} &
1 - \hspace{0.05cm} \cos^4\theta^{}_{14}
\left[\sin^2 2\theta^{}_{12} \cos^4\theta^{}_{13}
\sin^2 \frac{\Delta m^{2}_{21} L}{4 E} + \frac{1}{2}
\sin^2 2\theta^{}_{13} \left(\sin^2 \frac{\Delta m^{2}_{31} L}{4 E}
+ \sin^2 \frac{\Delta m^{2}_{32} L}{4 E}\right) \right.
\nonumber \\
& & \hspace{0.3cm} \left. + \hspace{0.1cm}
\frac{1}{2} \cos 2\theta^{}_{12} \sin^2 2\theta^{}_{13}
\sin \frac{\Delta m^{2}_{21} L}{4 E}
\sin \frac{\left(\Delta m^{2}_{31} + \Delta m^2_{32}\right) L}{4 E} \right]
\nonumber \\
& & \hspace{0.3cm} - \hspace{0.1cm} \sin^2 2\theta^{}_{14} \left[
\sin^2 \theta^{}_{13} \sin^2 \frac{\Delta m^{2}_{43} L}{4 E} + \frac{1}{2}
\cos^2 \theta^{}_{13} \left(\sin^2 \frac{\Delta m^{2}_{41} L}{4 E}
+ \sin^2 \frac{\Delta m^{2}_{42} L}{4 E}\right) \right.
\nonumber \\
& & \hspace{0.3cm} \left. + \hspace{0.1cm}
\frac{1}{2} \cos 2\theta^{}_{12} \cos^2 \theta^{}_{13}
\sin \frac{\Delta m^{2}_{21} L}{4 E}
\sin \frac{\left(\Delta m^{2}_{41} + \Delta m^2_{42}\right) L}{4 E} \right] \; ,
\end{eqnarray}
where the standard-like parametrization
$|U^{}_{e1}| = \cos\theta^{}_{12}\cos\theta^{}_{13}
\cos\theta^{}_{14}$, $|U^{}_{e2}| = \sin\theta^{}_{12}\cos\theta^{}_{13}
\cos\theta^{}_{14}$, $|U^{}_{e3}| = \sin\theta^{}_{13} \cos\theta^{}_{14}$
and $|U^{}_{e4}| = \sin\theta^{}_{14}$ \cite{Xing2011} have been adopted.
It is obvious that switching off the active-sterile neutrino mixing angle
$\theta^{}_{14}$ will allow Eq. (6) to reproduce the standard case shown
in Eq. (2). Some immediate comments on the new oscillatory terms
are in order.
\begin{itemize}
\item     The oscillatory term driven by $\Delta m^2_{43}$ is doubly suppressed
by the small flavor mixing factors $\sin^2 \theta^{}_{13}$ and
$\sin^2 2\theta^{}_{14}$, and hence it should not play an important role in
most cases, no matter how large or small the magnitude of $\Delta m^2_{43}$
could be.

\item     A sum of the two oscillatory terms driven by $\Delta m^2_{41}$
and $\Delta m^2_{42}$ has been extensively assumed to explain the so-called
LSND anomaly \cite{LSND}, the MiniBooNE anomaly \cite{MB} and the
reactor antineutrino anomaly \cite{RA} by requiring
$\Delta m^2_{4i} \sim {\cal O}(1) ~{\rm eV}^2$ (for $i=1,2,3$) \cite{Giunti}.
It remains unclear whether such anomalies are really a kind of signature of
possible new physics or just some kind of statistical or systematical
problems associated with the relevant measurements, but it is clear
that a light sterile antineutrino species with the ${\cal O}(1) ~{\rm eV}$
mass value cannot affect the main behaviors of $\overline{\nu}^{}_e
\to \overline{\nu}^{}_e$ oscillations in the JUNO experiment.

\item     One can see that the new interference term
is proportional to the product of
$\sin^2 2\theta^{}_{14}$, $\cos 2\theta^{}_{12}$, $\cos^2\theta^{}_{13}$
$\sin[\Delta m^2_{21} L/(4 E)]$ and
$\sin[(\Delta m^2_{41} + \Delta m^2_{42}) L/(4 E)]$, which is apparently
sensitive to the sign of $\Delta m^2_{41} + \Delta m^2_{42}$. Unless
the value of $m^{}_4$ is in between those of $m^{}_1$ and $m^{}_2$,
the mass-squared differences $\Delta m^2_{41}$ and $\Delta m^2_{42}$
should have the same sign. If we put together the standard and new
interference terms,
\begin{eqnarray}
{\rm Interference ~ terms} & \hspace{-0.2cm} =  \hspace{-0.2cm} &
\frac{1}{2} \cos 2\theta^{}_{12} \sin \frac{\Delta m^{2}_{21} L}{4 E}
\left[ \sin^2 2\theta^{}_{13} \cos^4\theta^{}_{14}
\sin \frac{\left(\Delta m^{2}_{31} + \Delta m^2_{32}\right) L}{4 E} \right.
\nonumber \\
& & \hspace{3.62cm} \left. +
\cos^2 \theta^{}_{13} \sin^2 2\theta^{}_{14}
\sin \frac{\left(\Delta m^{2}_{41} + \Delta m^2_{42}\right) L}{4 E} \right] \; ,
\hspace{1.25cm}
\end{eqnarray}
we might worry whether the latter would contaminate the former.
This is certainly dependent upon the possibly allowed ranges of
those new parameters (i.e., $\theta^{}_{14}$, $\Delta m^2_{41}$ and
$\Delta m^2_{42}$), and thus deserves a detailed analysis.
\end{itemize}
For the time being one has to admit that
the assumption of light sterile neutrino
species is primarily motivated at the phenomenology level and
lacks a strong theoretical motivation. In other words, it is unclear
why such light and sterile degrees of freedom should exist
and what place they could find in a more fundamental flavor theory
and (or) in the evolution of our Universe.

Zhou has pointed out that the probabilities of neutrino oscillations keep
invariant under the transformations $\theta^{}_{12} \to \theta^{}_{12} - 90^\circ$
and $m^{}_1 \leftrightarrow m^{}_2$ \cite{Zhou2017}. This intrinsic
symmetry can easily be seen in Eqs. (2) and (6), and it is also valid
in the presence of matter effects (i.e.,
$\widetilde{\cal P}(\overline{\nu}^{}_e \to \overline{\nu}^{}_e)$ keeps
invariant under the same transformations). Hence it looks quite natural
that the interference term under discussion should be proportional to the product of
$\cos 2\theta^{}_{12}$ and $\sin [\Delta m^2_{21} L/(4 E)]$, which
both change sign for $\theta^{}_{12} \to \theta^{}_{12} - 90^\circ$
and $m^{}_1 \leftrightarrow m^{}_2$. As a straightforward result,
the interference effect would vanish if $\theta^{}_{12} = 45^\circ$ held
\footnote{We would like to thank Shun Zhou for kindly pointing out this
interesting observation. It is mysterious that the neutrino sector seems
to only favor a kind of symmetry between the $\nu^{}_\mu$ and
$\nu^{}_\tau$ flavors (i.e., $\theta^{}_{23} \simeq 45^\circ$)
\cite{Zhao}.}.

\section{Discussions}

It makes sense to pursue a new formulation of the probability of
$\overline{\nu}^{}_e \to \overline{\nu}^{}_e$ oscillations
such that the interference effect, which can be used to probe
the neutrino mass ordering in the JUNO-like experiments, becomes
more transparent and understandable. We have done so by singling
out the interference term of
${\cal P}(\overline{\nu}^{}_e \to \overline{\nu}^{}_e)$ which is
proportional to the product of $\sin[\Delta m^2_{21} L/(4 E)]$ and
$\sin[(\Delta m^2_{31} + \Delta m^2_{32}) L/(4 E)]$. The same
interference term exists in the other disappearance-type neutrino
and antineutrino oscillations.

One may similarly express the probabilities of appearance neutrino and
antineutrino oscillations in terms of $\Delta m^2_{21}$ and
$\Delta m^2_{31} + \Delta m^2_{32}$, from which the interference
effects can be made transparent. We list the formula in the Appendix
for the sake of completeness. When the matter effects are
concerned for a long-baseline oscillation experiment,
a more general description in terms of an $\eta$-gauged
neutrino mass-squared difference
$\Delta^{}_* \equiv \eta \Delta m^2_{31} + \left( 1 - \eta\right)
\Delta m^2_{32}$ has been discussed by Li {\it et al} \cite{Li2017},
and the intrinsic symmetry of this description has been explored by
Zhou with the help of the renormalization-group language
\cite{Zhou2017}. Their findings indicate that the choice
$\eta = \cos^2\theta^{}_{12}$ seems to be most convenient in
making analytical approximations of
$\widetilde{\cal P}({\nu}^{}_\alpha \to {\nu}^{}_\beta)$
(for $\alpha, \beta = e, \mu, \tau$), but $\eta =1/2$ is certainly
an interesting option.

Is it necessary or useful to apply the same treatment to the formulation
of neutrino-antineutrino oscillations for the massive Majorana
neutrinos? The answer should be no, because the oscillation probabilities
${\cal P}(\nu^{}_\alpha \to \overline{\nu}^{}_\beta)$ (for
$\alpha, \beta = e, \mu, \tau$) directly depend on the neutrino
masses $m^{}_i$ (for $i=1,2,3$), the effective neutrino masses
$\langle m\rangle^{}_{\alpha\beta} \equiv
m^{}_1 U^{}_{\alpha 1} U^{}_{\beta 1} +
m^{}_2 U^{}_{\alpha 2} U^{}_{\beta 2} +
m^{}_3 U^{}_{\alpha 3} U^{}_{\beta 3}$ and
the neutrino mass-squared differences $\Delta m^2_{ji}$
(for $i, j =1,2,3$) \cite{XingZhou2013}. But such lepton-number-violating
processes are unfortunately suppressed by $m^{}_i m^{}_j/E^2$, and hence it is
actually hopeless to measure them in any realistic experiments
in the foreseeable future.

It is still a long way to verify or falsify the hypothetical
light sterile neutrinos, which have been introduced as a phenomenological
recipe to interpret some neutrino- or antineutrino-associated anomalies
at low energies. How to theoretically identify the physical significance
of such light ``dark" particles remains a big challenge today.

\vspace{0.5cm}

{\it The author would like to thank J. Cao, Y.F. Li, S. Zhou and J.Y. Zhu for
valuable discussions, and especially S. Zhou for reading the manuscript
and very helpful comments. This work was supported in
part by the National Natural Science Foundation of China under Grant
No. 11775231.}

\vspace{0.5cm}

\begin{center}
{\Large\bf Appendix}
\end{center}

In the standard three-flavor scheme the probabilities of 
$\nu^{}_\alpha \to \nu^{}_\beta$ oscillations (for
$\alpha, \beta = e, \mu, \tau$ but $\alpha \neq \beta$) can be expressed as 
\begin{eqnarray}
{\cal P}(\nu^{}_\alpha \to \nu^{}_\beta)
& \hspace{-0.2cm} =  \hspace{-0.2cm} &
- 2 \left(|U^{}_{\alpha 3}|^2 |U^{}_{\beta 3}|^2 -
|U^{}_{\alpha 1}|^2 |U^{}_{\beta 1}|^2 -
|U^{}_{\alpha 2}|^2 |U^{}_{\beta 2}|^2
\right) \sin^2 \frac{\Delta m^{2}_{21} L}{4 E}
\nonumber \\
& & - 2 |U^{}_{\alpha 3}|^2 \left(1 - |U^{}_{\beta 3}|^2\right)
\left[\sin^2 \frac{\Delta m^{2}_{31} L}{4 E}
+ \sin^2 \frac{\Delta m^{2}_{32} L}{4 E}\right]
\nonumber \\
& & + 2 \left(|U^{}_{\alpha 1}|^2 |U^{}_{\beta 1}|^2 - |U^{}_{\alpha 2}|^2
|U^{}_{\beta 2}|^2 \right) \sin \frac{\Delta m^{2}_{21} L}{4 E}
\sin \frac{\left(\Delta m^{2}_{31} + \Delta m^2_{32}\right) L}{4 E}  \hspace{0.6cm}
\nonumber \\
& & + 4 {\cal J} \epsilon^{}_{\alpha\beta\gamma}
\sin \frac{\Delta m^{2}_{21} L}{4 E}
\left[\cos \frac{\Delta m^{2}_{21} L}{4 E} -
\cos \frac{\left(\Delta m^{2}_{31} + \Delta m^2_{32}\right) L}{4 E}
\right] \; ,
\end{eqnarray}
where ${\cal J} = \sin 2\theta^{}_{12} \sin 2\theta^{}_{13} \cos\theta^{}_{13}
\sin 2\theta^{}_{23} \sin\delta/8$ is the well-known Jarlskog invariant 
\cite{J} which measures the strength of leptonic CP violation in neutrino 
oscillations, and the Greek subscripts $\alpha$, $\beta$ and $\gamma$ run
over the flavor indices $e$, $\mu$ and $\tau$. The expression of 
${\cal P}(\overline{\nu}^{}_\alpha \to \overline{\nu}^{}_\beta)$ can 
directly be read off from Eq. (8) with the replacement ${\cal J} \to
-{\cal J}$.


\begin{thebibliography}{99}

\bibitem{Reines}
C.L. Cowan, F. Reines, F.B. Harrison, H.W. Kruse, and A.D. McGuire,
Science {\bf 124}, 103 (1956).

\bibitem{KamLAND}
K. Eguchi {\it et al} (KamLAND Collaboration), Phys. Rev. Lett.
{\bf 90}, 021802 (2003).

\bibitem{DYB}
F.P. An {\it et al} (Daya Bay Collaboration), Phys. Rev. Lett.
{\bf 108}, 171803 (2012).

\bibitem{JUNO}
F. An {\it et al} (JUNO Collaboration), J. Phys. G {\bf 43}, 030401
(2016).

\bibitem{Wang}
L. Zhan, Y. Wang, J. Cao, and L. Wen, Phys. Rev. D {\bf 78},
111103 (2008); Phys. Rev. D {\bf 79}, 073007 (2009). See also:
J. Learned, S.T. Dye, S. Pakvasa, and R.C. Svoboda,
Phys. Rev. D {\bf 78}, 071302 (2008).

\bibitem{Lisi}
F. Capozzi, E. Lisi, A. Marrone, and A. Palazzo, arXiv:1804.09678;
P.F. de Salas, D.V. Forero, C.A. Ternes, M. Tortola, and J.W.F. Valle,
arXiv:1708.01186 (updated in April 2018);
I. Esteban, M.C. Gonzalez-Garcia, M. Maltoni, I. Martinez-Soler,
and T. Schwetz, JHEP {\bf 01} (2017) 087 [arXiv:1611.01514], and
NuFIT 3.2 (2018), www.nu-fit.org.

\bibitem{Li2013}
Y.F. Li, J. Cao, Y. Wang, and L. Zhan, Phys. Rev. D {\bf 88},
013008 (2013).

\bibitem{Petcov}
S.T. Petcov and M. Piai, Phys. Lett. B {\bf 533}, 94 (2002);
S. Choubey, S.T. Petcov, and M. Piai, Phys. Rev. D {\bf 68}, 113006 (2003).

\bibitem{Ge}
S.F. Ge, K. Hagiwara, N. Okamura, and Y. Takaesu, JHEP {\bf 1305},
131 (2013).

\bibitem{Ge2}
J. LoSecco, arXiv:1306.0845.

\bibitem{Cahn2013}
R.N. Cahn {\it et al.}, arXiv:1307.5487.

\bibitem{Lisi2014}
F. Capozzi, E. Lisi, and A. Marrone, Phys. Rev. D {\bf 89},
013001 (2014).

\bibitem{Wang2016}
Y. Wang and Z.Z. Xing, Adv. Ser. Direct. High Energy Phys. {\bf 26},
371 (2016).

\bibitem{Li2016}
Y.F. Li, Y. Wang, and Z.Z. Xing, Chin. Phys. C {\bf 40},
091001 (2016).

\bibitem{Zhu2018}
Y.F. Li, Z.Z. Xing, and J.Y. Zhu, Phys. Lett. B {\bf 782},
578 (2018).

\bibitem{Fogli}
G.L. Fogli, E. Lisi, A. Marrone, and A. Palazzo,
Prog. Part. Nucl. Phys. {\bf 57}, 742 (2006).

\bibitem{PMNS}
Z. Maki, M. Nakagawa, and S. Sakata, Prog. Theor. Phys. {\bf 28}, 870 (1962);
B. Pontecorvo, Sov. Phys. JETP {\bf 26}, 984 (1968)
[Zh. Eksp. Teor. Fiz. {\bf 53}, 1717 (1967)].

\bibitem{RENO}
J.K. Ahn {\it et al} (RENO Collaboration), Phys. Rev. Lett.
{\bf 108}, 191802 (2012).

\bibitem{DC}
Y. Abe {\it et al} (Double Chooz Collaboration), Phys. Rev. D
{\bf 86}, 052008 (2012).

\bibitem{MSW}
L. Wolfenstein, Phys. Rev. D {\bf 17}, 2369 (1978);
S.P. Mikheev and A.Y. Smirnov,
Sov. J. Nucl. Phys. {\bf 42}, 913 (1985).

\bibitem{Zhu2016}
Z.Z. Xing and J.Y. Zhu, JHEP {\bf 1607}, 011 (2016).

\bibitem{Xing2013}
Z.Z. Xing, Phys. Lett. B {\bf 718}, 1447 (2013).

\bibitem{Xing2011}
Z.Z. Xing, Phys. Lett. B {\bf 660}, 515 (2008);
Phys. Rev. D {\bf 85}, 013008 (2011).

\bibitem{LSND}
A. Aguilar {\it et al.} (LSND Collaboration), Phys. Rev. D
{\bf 64}, 112007 (2001).

\bibitem{MB}
A.A. Aguilar-Arevalo {\it et al.} (MiniBooNE Collaboration),
Phys. Rev. Lett. {\bf 105}, 181801 (2010).

\bibitem{RA}
G. Mention {\it et al.}, Phys. Rev. D {\bf 83}, 073006 (2011).

\bibitem{Giunti}
For recent reviews with extensive references, see:
S. Gariazzo, C. Giunti, M. Laveder, Y.F. Li, and E.M. Zavanin,
J. Phys. G {\bf 43}, 033001 (2016);
M. Drewes {\it et al.}, JCAP {\bf 1701}, 025 (2017).

\bibitem{Zhou2017}
S. Zhou, J. Phys. G {\bf 44}, 044006 (2017).

\bibitem{Zhao}
For a review with extensive references, see: 
Z.Z. Xing and Z.H. Zhao, Rept. Prog. Phys. {\bf 79}, 076201 (2016).

\bibitem{Li2017}
Y.F. Li, J. Zhang, S. Zhou, and J.Y. Zhu,
JHEP {\bf 1612}, 109 (2016).

\bibitem{XingZhou2013}
Z.Z. Xing, Phys. Rev. D {\bf 87}, 053019 (2013);
Z.Z. Xing and Y.L. Zhou, Phys. Rev. D {\bf 88}, 033002 (2013).

\bibitem{J} C. Jarlskog, Phys. Rev. Lett. {\bf 55}, 1039 (1985).

\end{thebibliography}
\end{document}